\documentclass{elsarticle}
\usepackage{amssymb}
\usepackage{pdflscape} 
\usepackage{multirow}
\usepackage{multicol}
\usepackage{appendix}
\usepackage{makecell}
\usepackage{hyperref}
\usepackage{pdfpages}
\usepackage{booktabs}
\usepackage{setspace}
\usepackage{float}
\usepackage{tabularx}
\usepackage{amsmath}
\usepackage{algorithm}
\usepackage{algorithmic}
\usepackage[normalem]{ulem}
\begin{document}
\begin{frontmatter}
\onehalfspacing
\title{Focused digital cohort selection from social media using the metric backbone of biomedical knowledge graphs}


\author[label1]{Ziqi Guo}
\author[label1]{Jack Felag}
\author[label1]{Jordan C. Rozum}
\author[label1]{Rion Brattig Correia}
\author[label3,label1]{Xuan Wang}
\author[label1,label2]{Luis M. Rocha}

\address[label1]{School of Systems Science \& Industrial Engineering, Binghamton University, Binghamton, NY, USA}
\address[label2]{Universidade Católica Portuguesa, Católica Biomedical Research Centre, Lisbon, Portugal.}
\address[label3]{School of Informatics, Computing \& Engineering, Indiana University, Bloomington, IN, USA.}
\ead{rocha@binghamton.edu}

\begin{abstract}

Social media data allows researchers to construct large \textit{digital cohorts}---groups of users who post health-related content—--to study the interplay between human behavior and medical treatment.
Identifying the users most relevant to a specific health problem is, however, a challenge in that social media sites vary in the generality of their discourse. While X (formerly Twitter), Instagram, and Facebook cater to wide ranging topics, Reddit subgroups and dedicated patient advocacy forums trade in much more specific, biomedically-relevant discourse.

To filter relevant users on any social media, we have developed a general method and tested it on epilepsy discourse.
We analyzed the text from posts by users who mention epilepsy drugs at least once in the general-purpose social media sites X and Instagram, the epilepsy-focused Reddit subgroup (r/Epilepsy), and the Epilepsy Foundation of America (EFA) forums.
We used a curated medical terminology dictionary to generate a knowledge graph (KG) from each social media site, whereby  
nodes represent terms, and edge weights denote the strength of association between pairs of terms in the collected text.

Our method is based on computing the metric backbone of each KG, which yields the (sparsified) subgraph of edges that participate in shortest paths. 
By comparing the subset of users who contribute to the backbone to the subset who do not, we show that epilepsy-focused social media users contribute to the KG backbone in much higher proportion than do general-purpose social media users.
Furthermore, using human annotation of Instagram posts, we demonstrate that users who do not contribute to the backbone are much more likely to use dictionary terms in a manner inconsistent with their biomedical meaning and are rightly excluded from the cohort of interest.

Our metric backbone approach, thus, has several benefits: 
it yields focused user cohorts who engage in discourse relevant to a targeted biomedical problem; 
unlike engagement-based approaches, it can retain low-engagement users who nonetheless contribute meaningful biomedical insights and filter out very vocal users who contribute no relevant content,
it is parameter-free, algebraically principled, does not require classifiers or human-curation, and is simple to compute with the open-source code we provide. 

\end{abstract}

\begin{graphicalabstract}
\includegraphics[width=1.0\textwidth]{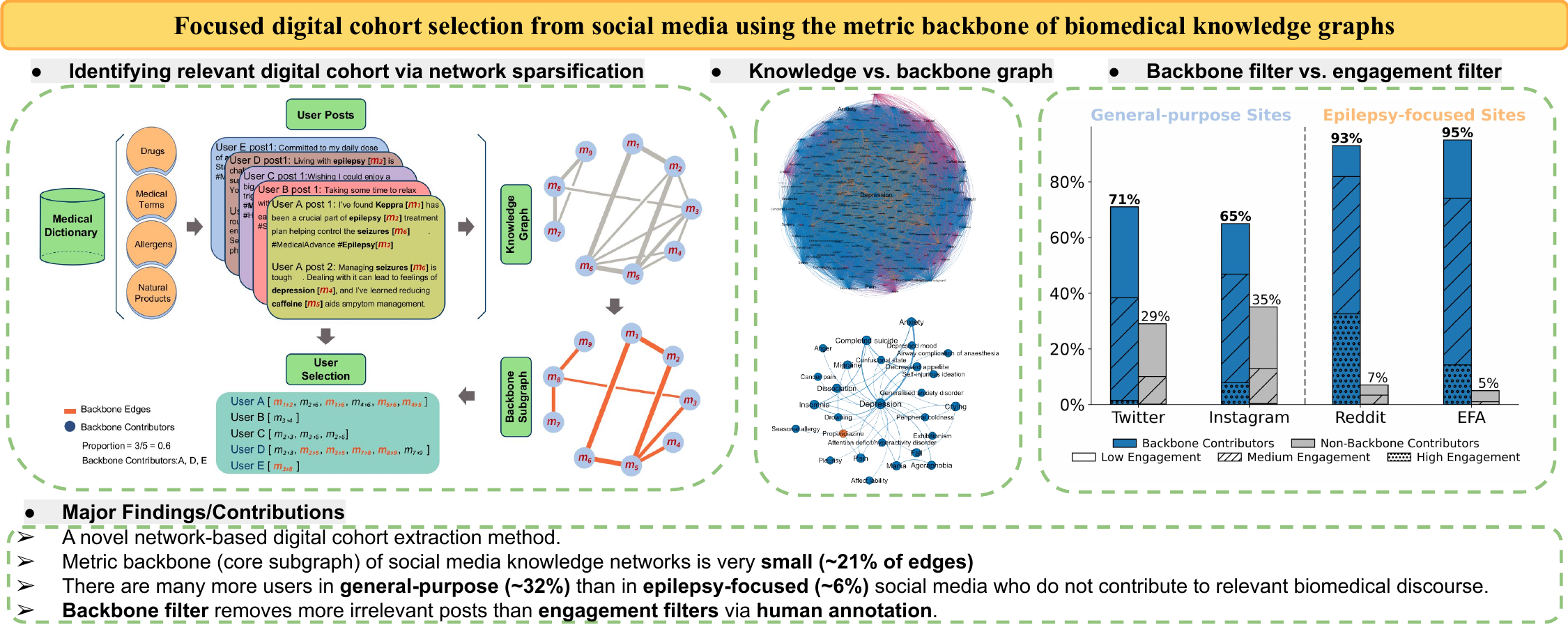}
\end{graphicalabstract}


\end{frontmatter}

\onehalfspacing
\section{Introduction} \label{intro}

Social media has gained significant attention in health research as a resource for building large \textit{digital cohorts}: an observed group of online users who engage with or post about specific health-related content \cite{correia2016monitoring,correia2020mining,wood2022small, gu_distilling_2023}.
Digital cohorts can supplement clinical trials and they provide several benefits.
First, they can capture information about patient-centered outcomes such as quality of life, satisfaction with treatment, and adherence to medication regimens \cite{correia2025myaura}. Second, they can be built from a more extensive population than clinical cohorts, which often have strict eligibility criteria that may exclude certain groups, such as children, patients with rare diseases or pregnant women \cite{wood2022small,meng2017social}. Third, digital cohorts can be constructed more expeditiously and less expensively than clinical cohorts because participant recruitment, medical testing, and longitudinal data collection plans are not required \cite{wood2017human}. Finally, monitoring medication and intervention-related adverse events in a digital cohort presents an opportunity to detect safety concerns, including misinformation and disinformation, at an earlier stage than with conventional pharmacovigilance techniques, which rely on reports from healthcare professionals \cite{correia2016monitoring,correia2019prediction,wang2020detecting}. These and other benefits of studying digital cohorts from social media data have been widely reported in the biomedical informatics literature \cite{correia2020mining,amir2019mental}.

Advances in computation, statistics, and mathematics have enabled automatic processing of data sets drawn from large digital cohorts using natural language processing, machine learning, and other data science methods. This has created new opportunities for researchers to tackle previously intractable problems that involve the interplay between human behavior and medical treatment. Examples include studying smoking cessation patterns using Facebook \cite{struik2014role}, monitoring potential drug interactions and reactions using Instagram \cite{correia2016monitoring}, detecting disease outbreaks early using X \cite{ofoghi2016towards}, analyzing the routes of administration and drug tampering for nonmedical opioid consumption using Reddit \cite{balsamo2021patterns}, and identifying putative predictors of Sudden Unexpected Death in Epilepsy using Facebook \cite{wood2022small}. Successful applications of social media data include biomedical signal detection and analysis of health-related behaviors \cite{correia2020mining}, pharmacovigilance \cite{sarker2015utilizing}, and public health disease surveillance \cite{aiello2020social, paul2016social}. 

Social media data is especially useful in investigating chronic conditions. It offers longitudinal insights into individual experiences and challenges over an extended duration, thereby providing valuable information on behaviors, habits, the exposome, psychology, and social interactions \cite{correia2020mining,de2024challenges}. Despite the numerous advantages of social media analysis in public health research, there has been limited focus on epilepsy, which is a chronic, noncommunicable brain disorder that ranks among the most common neurological conditions worldwide \cite{WHO2023}. Research suggests that among all neurological conditions, the number of online users interested in epilepsy is likely to be the highest \cite{meng2017social}. Thus, social media data related to epilepsy can provide insights into disease perception, patient needs, and access to treatments \cite{wood2022small, mcneil2012epilepsy, meng2017social}. However, to our knowledge, there have been no previous comparative analyses among different platforms in the context of epilepsy, or any chronic disease discourse, and prior research into the online behavior of people with epilepsy has primarily concentrated on isolated social media platforms. 

Social media site data are not equally suited to analysis for drawing biomedically relevant conclusions from patient discourse.
X, Facebook, and Instagram are general-purpose sites that cover a wide range of topics and only a small minority of posts pertain to epilepsy. In contrast, the Reddit subgroup (r/Epilepsy) and the Epilepsy Foundation of America (EFA) forums, and thus their posts, are dedicated to the discussion of epilepsy and related topics. 
Therefore, to focus on a given biomedical problem, filtering digital cohorts from platforms like X and Instagram is necessary to select user sets similar to those on special-purpose forums such as the EFA and r/Epilepsy.

Previous studies have identified cohorts of interest by, for example, matching certain keywords to posts on Instagram \cite{correia2016monitoring}, developing regular expressions to search for relevant users on X \cite{klein2022toward}, or considering all users who post content in a Subreddit related to a particular topic \cite{balsamo2021patterns}. However, data harvested in this way, especially from general-purpose social media sites, typically results in an overly broad cohort with a high volume of irrelevant posts. Additional filtering of users via engagement metrics is usually necessary to identify the most relevant ones, especially when the initial user pool is large. 
Unfortunately, engagement-based filtering (section \ref{sec:engagement method}) does not guarantee that selected users engage in discourse relevant to the specific topic of choice, especially when that topic is niche or obscure.
One way to address this limitation is to manually label posts to train classifiers for user filtering \cite{sarker2017discovering}. Such annotation efforts, however, are costly, laborious, and time-consuming. Therefore, an automated, general method to select relevant users for a topic of interest is highly desirable.

To address these issues, we present a novel network-based digital cohort selection method. The aim is to include only social media users who contribute to the shortest-paths of a knowledge graph (KG)---a network representation of entities and the relationship among them \cite{fensel2020introduction,correia2025myaura}---whose weighted edges represent measured associations between pairs of terms in a target dictionary pre-selected for biomedical relevance (section \ref{knowledge network}).
For our epilepsy study, we used a curated dictionary of medical terms from items extracted from Drugbank (v.5.1.0), MedDra (v15.0), MedlinePlus, and TCMGeneDIT (section \ref{dictionary}). It contains approximately 176k words that encompass drugs, allergens, medical terms, and natural products. This comprehensive dictionary comprises terms related not only to epilepsy but also to other diseases and conditions to facilitate the encoding of relevant associations \cite{correia2019prediction,min2024refinement,correia2025myaura}.

In a KG approach, the first assumption is that terms that co-occur in user posts more frequently than they appear with other terms share a strong association. 
In our method, the key additional assumption is that the semantic proximity between two terms in a KG is best captured by the strongest chain of associations, or the shortest path, between them.
Indeed, such shortest-path inference is very common in information retrieval and KG analysis as discussed in section \ref{knowledge network}. To identify which associations are required and which are redundant for the purpose of computing shortest paths (the strongest chains), we compute the \textit{metric backbone}
of each KG---an optimal sparsification method for weighted graphs such as KGs, detailed in section \ref{backbone}. 

The method we present and test here selects digital cohorts by identifying \textit{backbone contributors}: the users who co-mention (in at least one post) two terms linked by an edge of the metric backbone of a KG.
In other words, the users kept in our digital cohorts co-mention, at least once, two terms with a required association.
Users who are not backbone contributors are redundant in the sense that they only co-mention pairs of terms in our medical dictionary that are more strongly associated with other intermediary terms (i.e., they do not make any required associations for shortest paths).

Analysis of the digital cohorts thus identified, reveals that a much larger proportion of users contribute to the backbone in epilepsy-focused than in general-purpose social media. Specifically, only about $70\%$ of users initially selected from Instagram and X contribute to the metric backbone. In contrast, more than $90\%$ of users similarly selected from r/Epilepsy and the EFA forums contribute to the metric backbone (section \ref{result:different sites}).
As discussed below (section \ref{sec:discussion}), this suggests that in general-purpose social media sites, even among the users selected for mentioning epilepsy drugs, there is a substantial proportion who are redundant for biomedical inference, whereas in epilepsy-focused social media most such users are required for KG inference based on shortest paths.

We further investigate the biomedical relevance of retained and filtered-out users by using a corpus of Instagram posts (section \ref{Sec:InstagramAnnotatedSet}), which was previously human-annotated for potential medical relevance to epilepsy \cite{min2024refinement}. 
The proportion of users with false-positive dictionary term mentions (no medical relevance) is significantly higher among those who do not contribute to the metric backbone (32\%) than for those who do (14\%). 
This strongly suggests that the metric backbone approach is capable of filtering out irrelevant users, especially when compared to engagement-based filters, which do not (section \ref{compare filter methods}). 

In short, we show that the metric backbone of KGs is very effective in selecting focused digital cohorts with respect to a reference dictionary of relevant biomedical terms. Indeed, it allows the extraction of more biomedically-relevant cohorts from general-purpose social media sites (such as X) when more focused social media (such as community-specific forums) is not available.
\begin{table}[H]
    \centering
    \begin{tabular}{p{11.5cm}}\hline
         \textbf{Statement of Significance} \\ 
    \end{tabular}
    \begin{tabular}{p{3cm}p{8cm}}\hline
         \textbf{Problem:} & Identifying social media users and posts relevant to specific biomedical topics is challenging due to simultaneous discourse by many on a variety of topics on those platforms.\\ \hline
         \textbf{What is Already Known:} & Existing filtering methods identify relevant social media users according to metrics of engagement with terms of interest. However, quantity of matches does not guarantee relevance and can filter out relevant discourse about less frequent terms.\\ \hline
         \textbf{What this Paper Adds:} & A general-purpose network-based method for identifying relevant digital cohorts from social media, tested on epilepsy-relevant discourse. When applied to general-purpose social media platforms (X and Instagram), it yields digital cohorts akin to those obtained in platforms that focus on a specific biomedical topic (epilepsy subReddit and dedicated forums.) Furthermore, analysis of retained users reveals significantly higher post relevance.\\ \hline
         \textbf{Who would benefit from the new knowledge in this paper:} & Biomedical and social scientists, clinicians, technicians, and stakeholders interested in studying human behavior of biomedical relevance from social media data. \\ \hline
    \end{tabular}
    \label{tab:my_label}
\end{table}

\section{Data and methods}

\subsection{Medical dictionary curation} \label{dictionary}
We used a manually curated a medical dictionary to identify and aggregate terms relevant to the study of epilepsy. It was developed and curated in earlier research \cite{correia2016monitoring,zhang2022translational} using medical terms grouped in four categories: drugs, medical terminology, allergens and food items, and other natural products. The categories of terms are considered to encompass various factors that influence chronic disease management (e.g., dietary elements can impact treatment efficacy).
Terms can be associated with one or more of the four categories but these are not used in our analysis except for visualization purposes.
All terms were sourced from DrugBank (v.5.1.0) \cite{wishart2018drugbank}, MedDRA (v15.0) \cite{MedDRA}, MedlinePlus \cite{MedlinePlus}, and TCMGeneDIT \cite{TCMGeneDIT}. 
Dictionary terms are aggregated by mapping synonyms to a parent term. For instance, the epilepsy drug \emph{Levetiracetam} has the brand-name \emph{Keppra} as a synonym; \emph{Weed}, \emph{Mary Jane}, and \emph{420} are all synonyms of the parent term \emph{Cannabis}; the term \emph{Influenza} has \emph{Flu} and \emph{Flu syndrome} as synonyms. 
In addition, the dictionary was specifically tailored to social media discourse via human-annotation, whereby biomedical terms that have other senses in that context were removed\cite{min2024refinement}.
%
The resulting dictionary, used here, contains 145,229 terms, of which 75,023 are drugs, 66,956 are medical terms (including symptoms), 2,155 are allergens, and 1,173 are natural products (such as cannabis and food items).

\subsection{Social media data collection} \label{data collect}
We collected user-generated text related to epilepsy from four social media sites: X, Instagram, r/Epilepsy, and the EFA forums. An overview of the collected data is listed in Table \ref{tab:data_overview}. Here, ``digital cohort'' refers specifically to online social media users who are engaged with the topic of epilepsy per the retrieval and curation criteria detailed below for each platform. We should note that in other settings, this term may refer to the recruitment of participants for clinical research studies via social media.

\begin{table}[H]
\centering
\caption{Summary of collected social media data. Columns: \textbf{Raw digital cohorts} are the numbers of users harvested from the platforms before filtering. \textbf{Posts} are the numbers of posts from X and Instagram and posts and comments from r/Epilepsy and the EFA.}
\vspace{6pt}
\scriptsize
\label{tab:data_overview}
\begin{tabular}{ccccc}
\cline{2-5}
   & Platform & Collection Window & Raw digital cohorts & Posts \\ \hline
\multicolumn{1}{c}{\multirow{3}{*}{Drug Mention}} & X & 2011-2016 & 5,958 & 14,152,929 \\
\multicolumn{1}{c}{}                            & Instagram & 2010-2016 & 9,890 & 8,496,124 \\
\multicolumn{1}{c}{}                            & Reddit & 2018-2021 & 6,301 & 219,459 \\ 
\multicolumn{1}{c}{}                            & EFA & 2004-2016 & 8,488 & 78,948 \\ \hline
\end{tabular}
\end{table}

\subsubsection{General-purpose sites: X and Instagram} \label{sec:general-sites}
For the collection windows shown in Table A.1, we harvested all X and Instagram posts from users who mentioned ``\#seizuremeds" or at least one of seven drugs often used in the treatment of epilepsy; we refer to these users as \textit{Drug Mention} cohorts. We matched terms using whole-word matching and ignored hashtag symbols and case (e.g., ``seizuremeds'' and ``seizure meds'' are distinct, but ``Keppra'' and ``\#keppra'' are equivalent). A full list of drug terms used for matching is given in Supplementary Materials, Table A.1. This process resulted in over 14 million X posts from just under 6,000 users, and almost 8.5 million Instagram posts from just under 10 thousand users.

\subsubsection{Epilepsy-focused sites: Reddit(r/Epilepsy) and EFA}
\label{sec:focused sites}

Reddit users created the r/Epilepsy community in 2010 \cite{epilepsy} as a forum to support people affected by epilepsy. It includes those who have an epilepsy diagnosis, have seizures, are caregivers of people with epilepsy, and some advocates and healthcare professionals. Using the third-party API Pushshift \cite{baumgartner2020pushshift}, we collected all posts and comments on r/Epilepsy from June 1, 2018 (following changes to the Pushshift server) to December 31, 2021.

The EFA has granted us an exclusive use agreement to access data from their website (\url{epilepsy.com}), discussion groups, and social media presence. 
The EFA data set includes archived message boards, chat rooms, article comments, user symptom and medication diaries, and forums. In this work, we study archived forum posts between 2004 and 2016, which are organized by topic into several sub-forums (see Supplementary Table A.2).
 
To more directly compare the r/Epilepsy and EFA data sets to those from X and Instagram, we consider \textit{Drug Mention} cohorts within the selected r/Epilepsy and EFA users. These are the subsets of users who mention the epilepsy drugs in the same way as the selected X and Instagram users. In addition, we analyzed all posts and comments within our collection windows from r/Epilepsy and the EFA forum; these users form the \textit{All} cohorts, and the corresponding results can be found in Supplementary Material section E.

\subsubsection{Human-annotated Instagram Posts and False Positive Ratio}
\label{Sec:InstagramAnnotatedSet}

A random subset of Instagram posts was annotated with dictionary terms (Section \ref{dictionary}) on whether the mentioned dictionary terms were used in their expected sense, i.e., with putative medical relevance \cite{min2024refinement}. Through an extensive curation effort, 1,387 posts with 2,381 term occurrences, each annotated with consistent labels by two annotators, were included in this research. These annotated posts were sampled from 1,011 unique Instagram users.
If a term identified on Instagram matched the exact meaning in our biomedical dictionary within the context of that specific post, it is labeled a true positive, otherwise, it as a false positive. Because terms without medical meanings are not included in the dictionary, there are no negative instances, thus we do not have true negative or false negative labels. Please see the paper \cite{min2024refinement} for details of the annotation process and access to the data.

To assess how a digital cohort relates to biomedical discourse, we used the occurrences of annotated terms. Specifically, we calculated the false positive ratio (FPR), based on \textit{term-level} aggregation, as the number of false positive terms divided by the total number of annotated terms in a cohort. 
To compare two cohorts, we conducted a two-proportion $z$-test under the assumption that whether a term is false positive or true positive is independent and follows a binomial distribution, to measure the statistical difference. 
In addition, we provided an alternative approach, in Supplementary Material section I, to access biomedical relevance of a given cohort by calculating the FPR based on a lenient \textit{user-level} aggregation. Users with at least one true positive term match, are labeled true positive users, and otherwise, false positive users. The FPR is then determined as the proportion of users labeled as false positive.


\subsection{Engagement-based filtering} \label{sec:engagement method}

Engagement-based filtering methods are commonly used in the literature. They include filtering according to engagement metrics such as posting activity above a specified level, posting in a specified language at least a minimum percentage of the time \cite{syarif2019study}, and posting items valued by others, as evidenced by likes and retweets \cite{diaz2022noface,kumari2022social}. For this study, we employ user engagement statistics---including activity duration, number of posts per user, number of words per user, and dictionary term matches per user---to implement both a lenient and an aggressive set of filtering criteria. 
The lenient engagement filter selects users exceeding the 25th percentile in activity duration, post count, and word count among all users, and whose posts contain at least two unique medical terms. The aggressive filter raises the threshold to the 75th percentile but maintains the requirement for two unique medical terms. 
Before applying these filters, we excluded from the EFA analysis five users who we manually identified as having substantially higher levels of engagement (approximately 100 times) than the user average{---these five users are Epilepsy professionals associated with EFA who often respond to patient questions}.
We do so to prevent their disproportionately high levels of input from dominating the KG. In essence, we place greater value on contributions from typical users and seek to avoid biases introduced by outlier users.
Similarly, we exclude the \textsc{AutoModerator} bot account from the r/Epilepsy data set and any posts from subsequently deleted accounts (in these cases, the post author is labeled as \textsc{[deleted]}). We did not identify individual users who warranted manual removal in the data sets from X and Instagram.

Table A.6 in Supplementary Material displays the number of users retained after each of the lenient and aggressive filtering techniques, along with corresponding engagement statistics within the sites. Roughly half of the users were retained after lenient filtering and approximately 10\% were retained after aggressive filtering for all four social media sites.

\subsection{Building the knowledge graphs} 
\label{knowledge network}
To construct a KG for each social media data set considered here, we compute the Jaccard similarity co-occurrence score between each pair of dictionary terms that occurs in the data set. 
Co-occurence scores are determined with respect to a specified time window in each user's post history. Here, we compute co-occurence scores at the level of individual posts, but our approach generalizes to longer time windows in a straightforward way\cite{correia2016monitoring,correia2025myaura}. 
From this pair-wise score, we construct an edge-weighted, undirected proximity graph, where nodes represent terms and edges represent strength of co-occurrence. In this proximity graph, an edge weight close to one indicates that the linked terms nearly always appear together (are highly proximal), while an edge weight close to zero indicates that the terms usually appear separately. We omit edges that correspond to pairs of terms with fewer than three co-occurrences due to the high sensitivity to individual posts. 
To ensure enough support exists in the data for proximity associations, we computed proximity weights only when the total occurrences of two terms minus their co-occurrences is larger than ten, otherwise, we set proximity weights equal to 0. 
Thus, the proximity graph encodes clusters of terms that are highly associated within the discourse of the online community from which the graph is built.

These proximity graphs are simple, yet powerful, data representations for studying the associations among different terms, which are commonly used in problems such as recommender algorithms \cite{simas2015distance}, scientific maps \cite{manz2022viv,Borner:2021}, and many biomedical inference applications \cite{Abi-Haidar:2008,Verspoor:2005,Kolchinsky:2010,sanchez2024prevalence,correia2016monitoring}.
Many of the network inference methods used to analyze this type of KG depend on computing shortest paths between nodes \cite{Borner:2007:Net, Wasserman:1994,Monge:2003, Barabasi:2003, correia2025myaura}. 
From a KG of terms, we can infer indirect term associations: If $i$ is associated with $k$, and $j$ is also associated with $k$, the indirect path quantifies how close $i$ is to $j$ via $k$.
This type of inference is ubiquitous in KG problems such as link prediction \cite{LuLinyuan:2011}, recommendation \cite{LiXin:2013, DongYu:2012}, automated fact-checking  \cite{ciampaglia2015computational}, and network information propagation in general \cite{correia2023contact,soriano2023semi}---including in biomedical relation extraction problems such as protein-protein \cite{krallinger2011protein,abi2008uncovering} and drug-drug interactions \cite{correia2016monitoring,correia2019prediction,sanchez2024prevalence}.

Note that to compute shortest paths (to infer how terms may be related through indirect associations), we need to transform edge weights from proximity to distance. The proximity score (Jaccard similarity) $p_{ij}$ is the ratio of the number of posts that contain both terms $i$ and $j$ to the number of posts that contain either or both terms. We transform it to distance using $d_{ij} = 1/p_{ij} - 1$, as described in previous research \cite{simas2015distance, rocha2005mylibrary}. Thus, maximally proximal terms ($p_{ij}=1$) are mapped to a distance of $d_{ij}=0$, while minimally proximal terms ($p_{ij}=0$) are mapped to an infinite distance. 
By working in this distance space, we can use path lengths to explore the strength of direct and indirect associations between terms. Shortest paths in the KG represent chains of strong associations between terms in the online community under study.
For instance, in Figure \ref{fig:example_egoNet}.\textbf{c}, the term ``Migraine'' is associated with ``Depression'' ($d=30$) and with ``Topiramate'' ($d=52$), thus, the length of the indirect path from ``Depression'' via ``Migraine'' to ``Topiramate'' is $d=30+52=82$. Interestingly, this indirect path is shorter than the direct association between ``Depression'' and ``Topiramate'' ($d=1021$); it is the shortest path, or strongest association chain, between those terms, which makes sense since ``Topiramate'' is prescribed to prevent Migraine. The fact that frequently indirect paths are shorter than direct associations is key to our method, as we show in Section \ref{sec:results}.
Because the proximity-to-distance map is an isomorphism \cite{simas2015distance}, it is always possible to freely transition between the distance and proximity graphs.

In this way, we build biomedically-relevant (proximity and isomorphic distance) KGs from each of four social media platforms. Taken together, the users whose posts contribute to at least one edge in any of the KGs form the initial digital cohort, shown in Table \ref{tab:full_users}. Approximately 95\% of users whose posts were collected are retained in the initial digital cohort. Similarly, approximately 99\% of posts are retained.

\begin{table}[H]
\centering
\caption{KG statistics for each social media platform. \textit{Full digital cohort} is the number of users whose posts contribute to at least one edge of the KG. The ratio (\%) denotes the proportion of users or posts of the raw digital cohorts (table \ref{tab:data_overview}) that remain in the full digital cohorts.}
\vspace{6pt}
\scriptsize
\label{tab:full_users}
\begin{tabular}{cccccc}
\cline{2-6}
   & Platform & Nodes & Edges & Full digital cohorts (\%) & Posts (\%) \\ \hline
\multicolumn{1}{c}{\multirow{3}{*}{Drug Mention}} & X & 1,022 & 5,082 & 5,933 (99.6\%) & 14,143,622 (99.9\%)\\
\multicolumn{1}{c}{}                            & Instagram & 1,686 & 25,235 & 9,604 (97.1\%) & 8,418,628 (99.1\%)\\
\multicolumn{1}{c}{}                            & Reddit & 1,270 & 17,558 & 6,153 (97.7\%) & 218,342 (99.5\%)\\ 
\multicolumn{1}{c}{}                            & EFA & 1,529 & 33,795 & 8,353 (98.4\%) & 78,768 (99.8\%)\\ \hline
\end{tabular}
\end{table}

\section{Results}
\label{sec:results}

To demonstrate our digital cohort selection method, we
%
first define and compute the metric backbones of the KGs derived the four social media platforms (Section \ref{backbone}) and show that they are much sparser than their original KGs (Section \ref{sec:result_backbone}). This demonstrates that KGs obtained from biomedically-relevant social media data contain a large proportion of redundant edges for shortest path inference.
Based on that experimental observation, we then introduce the novel backbone-based filtering method for identifying biomedically relevant digital cohorts (Section \ref{methodological result}). Next we use that method to compute and analyze epilepsy-relevant digital cohorts, showing that it filters out many more users in the general-purpose social media platforms than in the epilepsy-focused ones (Section \ref{result:different sites}). 
Finally, using the corpus of annotated Instagram posts (Section \ref{Sec:InstagramAnnotatedSet}), we show that our method outperforms engagement filters in identifying the most relevant biomedical discourse and users (Section \ref{compare filter methods}).

\subsection{Metric backbone of the social media knowledge graphs}
\label{backbone}

As discussed in the Section \ref{knowledge network}, many KG inference methods depend on computing shortest paths.
Interestingly, many of the edges in KGs are completely redundant for such information retrieval and inference \cite{simas2015distance,simas2021distance} because they do not participate in any shortest path. 
These redundant edges can obscure key features of the network (e.g., in visualization and clustering) and reduce computational efficiency \cite{correia2023contact,correia2025myaura}.
For instance, 98\% of the edges of a KG of over 3 million terms extracted from Wikipedia are redundant for shortest-path inference, and their removal facilitates automated fact-checking \cite{ciampaglia2015computational,simas2021distance}.

In previous work \cite{simas2015distance,simas2021distance}, we developed a general framework to sparsify any weighted graph (such as the KGs considered here) to its \textit{metric backbone}: a subgraph that is sufficient to compute every shortest path of the original graph.
The length of path (a sequence of edges) from $i$ to $j$ via $k$ is $\ell \{i,k,j\} = d_{ik} + d_{kj}$.
As shown in Simas et al. \cite{simas2021distance}, edges $d_{ij}$ of distance graph $D$ that satisfy the triangle inequality $d_{ij} \leq d^C_{ik} + d^C_{kj} \; \forall k$ are necessary and sufficient to compute all shortest paths of $D$, 
%
while those that break this inequality are \textit{redundant} for this purpose. Note that $d^C_{ij}$ denotes the length of the shortest (direct or indirect) path between nodes $i$ and $j$ in the graph (see example below and previous work \cite{simas2021distance,correia2023contact} for details).
Thus, for a distance graph $D$, the \textit{metric backbone} $B$ is the subgraph composed of the edges that obey the triangle inequality. Broadly, generalized triangle inequalities for other measures of path length $\ell$ yield other distance backbones \cite{simas2021distance}. Here, we only use the metric backbone with $\ell \{i,k,j\} = d_{ik} + d_{kj}$. Therefore, henceforth, when we use the term backbone alone, we mean the metric backbone.

The metric backbone is an algebraically principled network sparsification method with unique features: it (a) preserves all connectivity and shortest paths, (b) does not alter edge weights or delete nodes, (c) is exact, not sampled or estimated, and (d) requires no parameters or null model estimation \cite{simas2021distance}. Furthermore, it outperforms others in (e) preserving the community structure of the original graph and (f) recovering most of the original (macro and micro) transmission dynamics in social contact networks, while revealing the most important infection pathways in epidemics and resulting in greater reduction without breaking apart the original network \cite{correia2023contact,soriano2023semi}.

It is easy to construct the metric backbone by computing the shortest indirect distance $d^C_{ij}$ between each pair of connected nodes $i$ and $j$, and comparing this value to the distance edge weight $d_{ij}$ connecting $i$ and $j$ directly. An edge $d_{ij}$ appears in $B$ if and only if it is an edge of $D$ with $d_{ij}=d^C_{ij}$. We use the Python package \textsc{distanceclosure} \cite{simas2021distance} to compute the metric backbone of all KGs. The package is publicly available via our lab's GitHub\footnote{ https://github.com/CASCI-lab/distanceclosure}. The implementation we use relies on the Dijkstra algorithm for shortest-path computation, and the worst case scaling of the backbone computation is $\mathcal{O}(N^3)$, where $N$ is the number of nodes in the KG.

\begin{figure}
    \centering
    \includegraphics[width=1\textwidth]{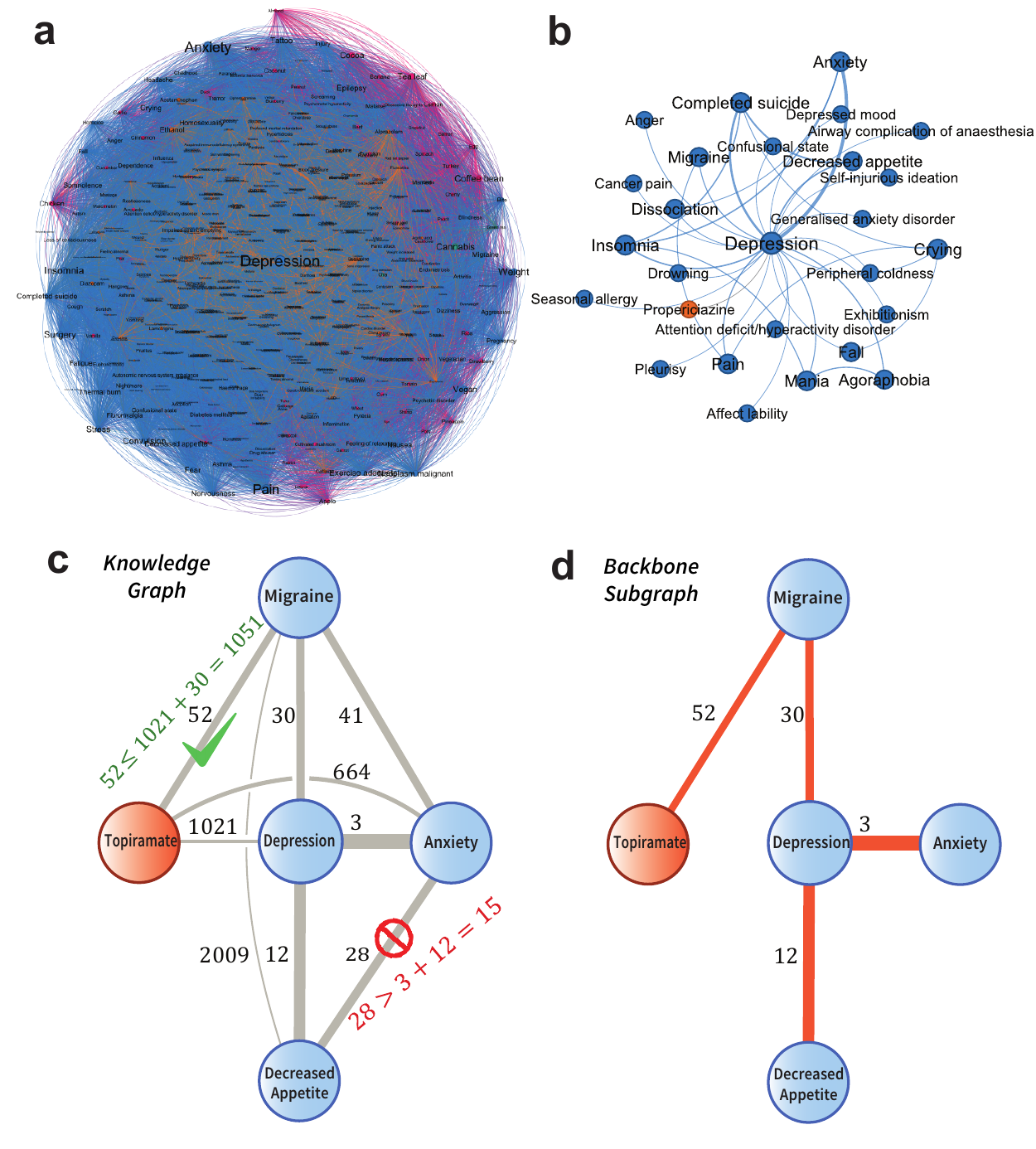}
	\caption{Ego network reduction using the metric backbone of the Instagram KG 
    \textbf{a}: The full ego network for the term ``Depression'' of the Instagram KG, a condition highly co-morbid with epilepsy, that shows the subnetwork of nodes directly connected to ``Depression'' in the Instagram KG. Rendered using a ForceAtlas2 \cite{jacomy2014forceatlas2} layout. 
    \textbf{b}: The metric backbone of the ``Depression'' ego network of the Instagram KG.
    \textbf{c}: Small subgraph of full ``Depression'' ego network shown in \textbf{a}, with five dictionary terms. Distance edge weights are obtained from term co-occurrence in posts (see Section \ref{knowledge network} for calculation details), with thickness rendered inverse proportionally to distance. 
    \textbf{d}: The metric backbone of the network in panel \textbf{c}; see text for additional details and examples.}

	\label{fig:example_egoNet}
\end{figure}

Subgraphs of the metric backbone of the Instagram KG are shown in Figure \ref{fig:example_egoNet}. 
%
Specifically, panel \textbf{a} shows the (full) ego network of this KG centered on the node for term ``Depression''.
An ego network (short for ego-centric network), consists of a focal node (the ``ego'') and the nodes directly connected to it (the ``alters''), plus any connections among the alters.
In this case, the ego node is the dictionary term ``Depression''. Thus edges represent the co-occurrences of this term with other dictionary terms in the Instagram posts of the full digital cohort. Panel \textbf{c} shows a small subgraph of this ego network selected to exemplify the metric backbone method used to produce the graphs in panels \textbf{b} and \textbf{c}.
In the original KG (Figure \ref{fig:example_egoNet}\textbf{a},\textbf{c}), numerous paths connect nodes, including redundant edges. The metric backbone greatly sparsifies the KG by retaining only the edges that obey the triangle inequality, which are necessary and sufficient to preserve all shortest paths as shown in Figure \ref{fig:example_egoNet}\textbf{d}. 
For instance, using the ``Topiramate'' to ``Depression''example presented in section \ref{knowledge network}, the direct edge between these terms ($d=1021$) breaks the triangle inequality ($d=1021 >= 52+30 = 82$), it is thus redundant for shortest paths and removed.

Notice that the metric backbone sparsification is quite different from thresholding the KG \cite{simas2021distance,correia2023contact}. For instance, distance edges between ``Decreased Appetite'' and ``Anxiety'' ($d=28$) and between ``Migraine'' and ``Anxiety'' ($d=41$) in Figure \ref{fig:example_egoNet}\textbf{c} are removed in the metric backbone in Figure \ref{fig:example_egoNet}\textbf{d} , but the distance edge between ``Migraine'' and ``Topiramate'' ($d=52$) is not, even though the latter is larger.

\subsection{The metric backbone of social media knowledge graphs is very small}
\label{sec:result_backbone}

\begin{figure}[H]
    \centering
	\includegraphics[width=1\textwidth]{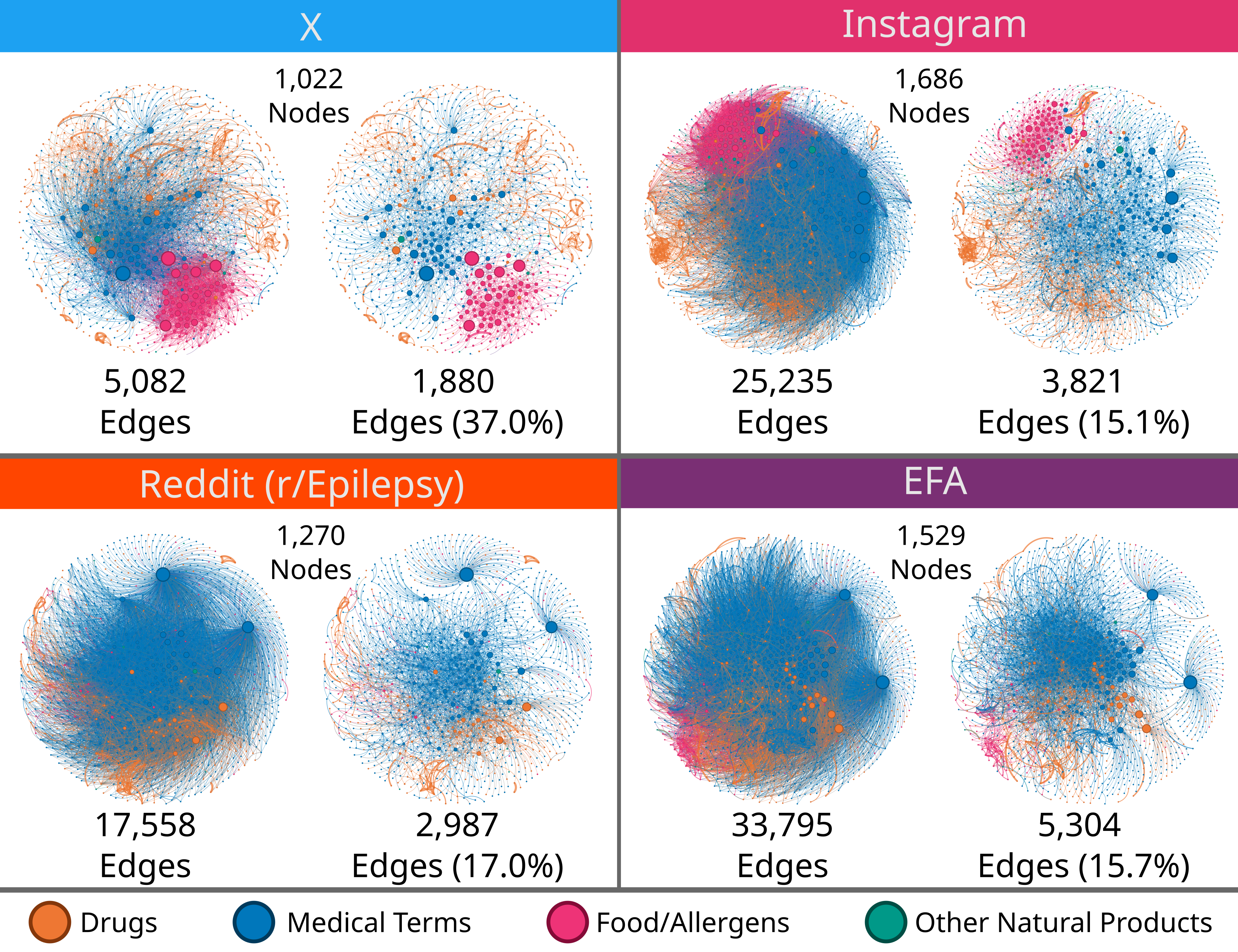}
	\caption{Social media KGs and their metric backbones. For each data source, the depiction on the left is the whole network, and on the right is the metric backbone subgraph. The relative size of the metric backbone is shown as the percentage of edges kept (all nodes are kept in the metric backbone). Dictionary terms can be associated with one or more of the four categories: drugs, medical terms, food/allergens, and other natural products as described in Section \ref{dictionary} with the colors shown in legend. These categories are used primarily for visualization purposes and do not impact our analysis. In case a term belongs to one or more classes, it is assigned to only one class in the network visualizations with the following preference order: drugs, medical terms, food/allergens, natural products.
    The nodes are sized according to their (unweighted) degree in the original network. The r/Epilepsy and EFA networks are computed only from their drug-mention subcohorts to better compare them with the X and Instagram networks. Node positions are determined by the Fruchterman Reingold method applied to the backbone networks.}
	\label{fig:NetworkOverview}
\end{figure}

We construct KGs from the four social media platforms and compute their metric backbones, as detailed in Sections \ref{knowledge network} and \ref{backbone}. 
Figure \ref{fig:NetworkOverview} depicts those KGs, their metric backbones, and the proportion of edges retained after this sparsification. It is clear that the metric backbones of all platforms are considerably smaller than the original KGs, indicating substantial redundancy. In other words, the original KGs can be substantially reduced, while still maintaining the associations required to preserve all shortest paths. Specifically, the metric backbones of the KGs from X, Instagram, r/Epilepsy, and the EFA forums are similar in size: $\approx$16\% of their original KGs. The metric backbone of the X KG is larger at 37\% of its original KG. The observed redundancy is similar to what has previously been reported for KG metric backbones \cite{simas2021distance}.

The extensive sparsification uncovered by the metric backbone also allows us to focus on specific discourse of interest. For instance, it is known that there is a strong comorbidity between epilepsy and depression \cite{fryska2023co, lotfaliany2018depression}. We can study the discourse surrounding depression within epilepsy communities on social media using, for instance,  the ego networks centered on the dictionary term ``Depression'' shown in Figure \ref{fig:example_egoNet}, which were introduced in section \ref{backbone}.

The metric backbone of the  full ``Depression'' ego network (Figure \ref{fig:example_egoNet}\textbf{b}) is much sparser than the original KG (panel \textbf{a}). Indeed, most nodes are no longer directly connected to ``Depression'', because many indirect associations are stronger than the direct associations in the KG. For example,``Topiramate'' is no longer directly connected to ``Depression'' in the metric backbone and is thus not shown in panel \textbf{b}.
Importantly, reachability is fully preserved by the metric backbone as the removed edges are redundant for shortest paths, e.g., the shortest distance between ``Topiramate'' and ``Depression'' is preserved, as are all other shortest paths.
Indeed, all nodes shown in Figure \ref{fig:example_egoNet}\textbf{a} are reachable from the ``Depression'' node via an indirect path on the full backbone KG (Figure \ref{fig:NetworkOverview}) with the same shortest distance as in the original KG.
However, most are no longer directly connected to ``Depression'' in the metric backbone, as shown in Figure \ref{fig:example_egoNet}\textbf{b}. 
Importantly, the nodes that remain directly connected to ``Depression'' in the metric backbone, satisfy the transitivity criterion of the triangle inequality defined in Section \ref{backbone}: the direct distance from each remaining node to ``Depression'' is never larger than the length of an indirect path to this target.
Similarly, many edges between nodes disappear in the ``Depression'' ego-backbone, because their relationship via intermediary terms is stronger than it is via direct co-occurrence.

In summary, by focusing on the metric backbone, we maintain only the key relationships that contribute to the shortest paths, simplify analysis, reduce computational complexity, and enhance visualization clarity while retaining essential connections for accurate inference.
In our example, the dictionary terms that remain connected in the ``Depression' ego-backbone are, from the perspective of shortest-paths, the most relevant to understand how the Epilepsy digital cohort extracted from Instagram discusses this term.
In a more general biomedical context---such as identifying treatment effects or explaining correlations and even causal relationships---the backbone subgraph of KGs helps focus on the most informative connections without redundant details \cite{correia2025myaura}.

\subsection{Focused digital cohort selection via the metric backbone of KGs}
\label{methodological result}

To select a biomedically-relevant digital cohort from social media data, we present a novel filtering method. It leverages KG redundancy to reveal which users engage with the central knowledge structure of a community, as characterized by the  \emph{metric backbone} of KGs (Section \ref{backbone}). Specifically, we identify the subset of users whose posts contribute to at least one edge on the metric backbone. We refer to these users as \emph{backbone contributors} throughout this paper. In other words, to obtain the set of backbone contributors, we exclude users whose posts do not contain the co-occurrence of a pair of terms associated with least one edge on the metric backbone. 

\begin{figure}
    \centering
	\includegraphics[width=1\textwidth]{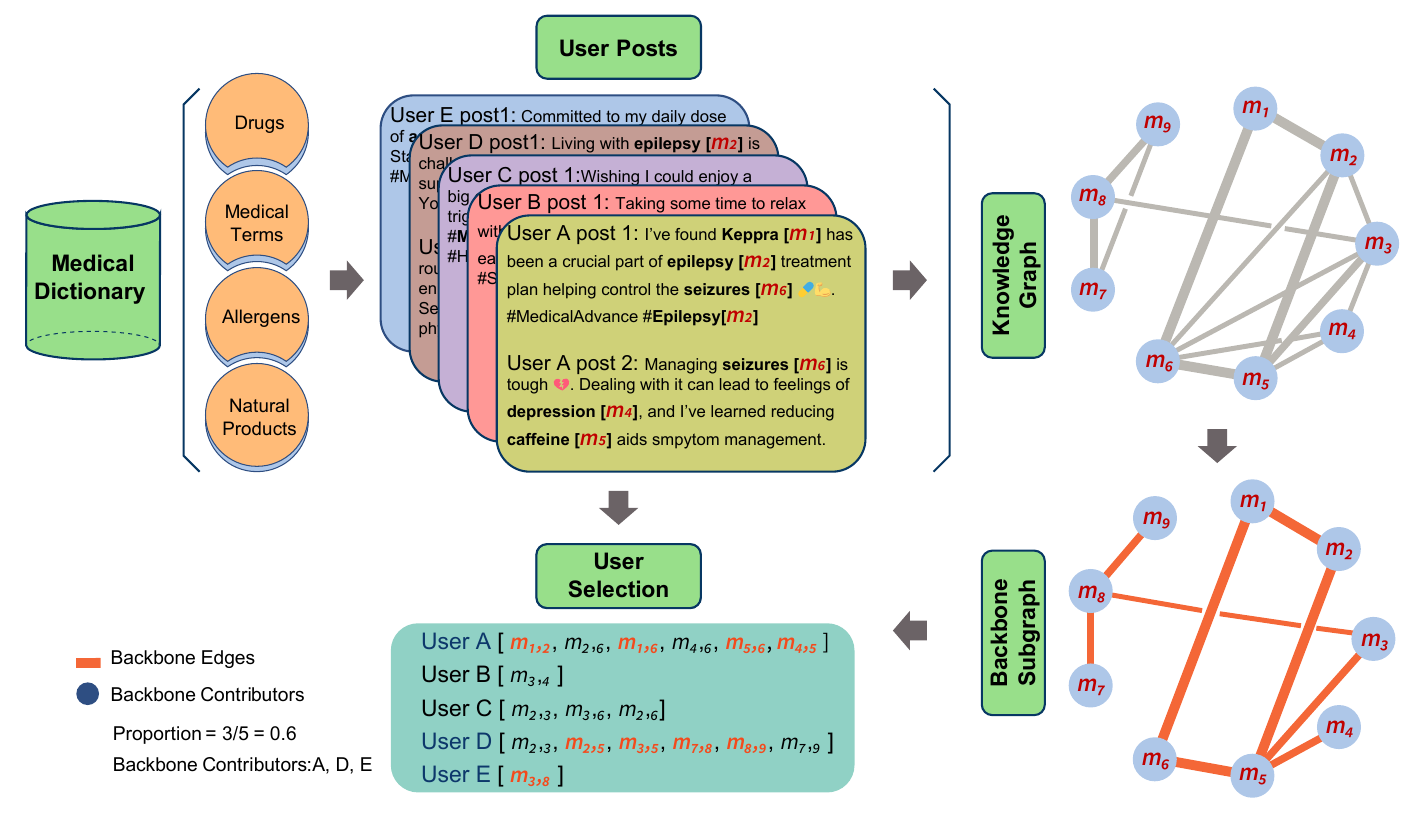}
	\caption{A backbone-based filtering example. First, we curate a medical dictionary, which includes terms related to drugs, allergens, medical terms, and natural products (see section \ref{dictionary}). Then, we collect posts from users on social media and match these posts with the dictionary terms (see section \ref{data collect}). The matched terms are represented by $m$ in the figure. Second, we build the KG, wherein the nodes represent the medical dictionary terms and the edge weights denote the likelihood that the connected pair of terms occur within same post (see section \ref{knowledge network}). Third, from the KG we compute the metric backbone (see section \ref{backbone}). Finally, we identify the \emph{backbone contributors}.}
	\label{fig:method schema}
\end{figure}

Figure \ref{fig:method schema} illustrates the overall workflow. The pipeline begins with tagging social media posts with the medical dictionary and ends with identifying the backbone contributors. In the example, there are five users (labeled A through E) in total; among them, three (A, D, and E) are backbone contributors, constituting 60\% of the total in this example. 
Our hypothesis is that a higher proportion of backbone contributors indicates a more focused discourse with respect to the chosen dictionary. In other words, if a cohort of social media users is actively engaged with a target biomedical terminology (as captured by a dictionary), we expect them to use the key associations concurrently as term co-occurrences (here, in social media posts). Our assumption is that key associations are those that form shortest paths in the KG and thus captured by the metric backbone. The results presented in sections \ref{result:different sites} and \ref{compare filter methods} below strongly support the hypothesis and assumption. See the pseudo-code for the proposed digital cohort filtering method based on the metric backbone in Supplementary Material section A.

\subsection{Epilepsy-focused sites have a higher proportion of backbone contributors}

\label{result:different sites}

When we apply our focused digital cohort selection method (Section \ref{methodological result}) to two general-purpose sites, X and Instagram, and two epilepsy-focused sites, r/Epilepsy and the EFA forums, the results indicate a significant disparity between the two sets of sites in terms of user relevancy, even though we harvested their user timelines with the same criterion: at least one post with an epilepsy drug mention. 
As shown in Table \ref{table: backbone contributors}, a much higher proportion of users contribute to the backbone ($r_{\text{raw}}$) in epilepsy-focused than in general-purpose social media: $\approx 93\%$ and $\approx 95\%$ on r/Epilepsy and EFA, versus $\approx 65\%$ and $\approx 71\%$ on Instagram and X, respectively. 
This shows that in the general-purpose platforms there are a lot more users who only contribute to redundant KG edges---term co-occurrences that do not participate in shortest paths on KGs, and thus not needed for such inference.
In other words, our results support the hypothesis that in social media platforms that are focused on a specific biomedical topic (epilepsy in our study), more users contribute to relevant biomedical terminology and thus to the metric backbone of derived KGs. Conversely, as we show next, when we extract cohorts from general-purpose social media platforms, even when focusing on users who mentioned drugs of interest, we likely harvest users whose discourse is irrelevant to the biomedical focus.
%

\begin{table}[H]
\centering
\caption{The proportion of backbone contributors from general-purpose and epilepsy-focused social media. $r_{\text{raw}}$ denotes the percentage of the backbone contributors relative to the raw (full) digital cohort.}
\scriptsize
\label{table: backbone contributors}
\begin{tabular}{cccc}
\cline{2-4}
   & Platform & Backbone contributors ($r_{raw}$) & Posts (\%)\\ \hline
\multicolumn{1}{c}{\multirow{3}{*}{Drug Mention}} & X & 4,219 (\textbf{70.8\%}) & 10,412,559 (73.6\%)  \\
\multicolumn{1}{c}{}                            & Instagram & 6,422 (\textbf{64.9\%}) & 7,118,578 (83.79\%)\\
\multicolumn{1}{c}{}                            & Reddit & 5,881 (\textbf{93.3\%}) & 217,718 (99.2\%)\\ 
\multicolumn{1}{c}{}                            & EFA & 8,076 (\textbf{95.1\%}) & 78,296 (99.2\%)\\ \hline
\end{tabular}
\end{table}


\subsection{Backbone-based method filters out users with incorrect term use better than engagement metrics}

\label{compare filter methods}

\begin{figure}[H]
    \centering
	\includegraphics[width=1\linewidth]{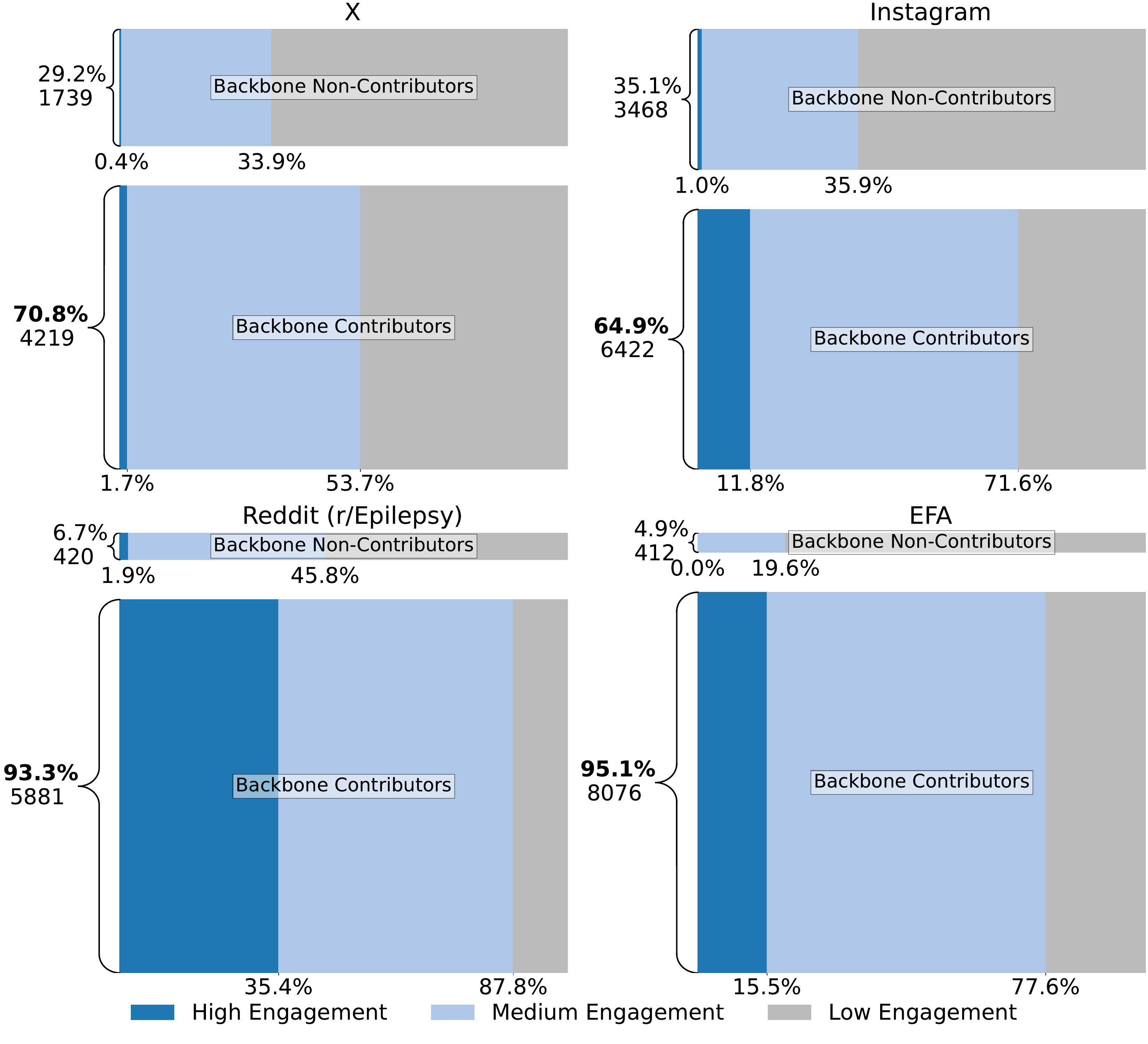}
	\caption{Proportions of users retained by backbone and engagement filters. High, medium, and low engagement refer to the filtering criteria of aggressive, lenient, and no filtering, respectively. Horizontal axes denote derived cohorts according to engagement filtering, and vertical axes denote derived cohorts according to backbone filtering. Each subset represents, for example, in the top left panel, 70.8\% (4,219) of users in the X cohort are Backbone Contributors, 1.7\% ($4,219 \times 1.7\% = 72$) of whom are high-engagement as well. The raw count of each subset can be found in Figure .6 in Supplementary Material.}
	\label{fig:overlap users}
\end{figure}

We compare our proposed method with the engagement-based filtering approach and validate the effectiveness of the proposed method using a human-annotated corpus of Instagram posts. The overlap in users retained by these filters across four social media sites is shown in Figure \ref{fig:overlap users}. In all four cohorts, backbone contributors are more likely to be high- and medium-engagement users than the backbone non-contributors, but a significant portion of low-engagement users do contribute to the metric backbone. This indicates that some term associations that are essential to compute the shortest chains of associations between medically relevant terms are made by low-engagement users.
In the most extreme case (X), nearly $50\%$ more users are retained by the backbone-based filter than by even the lenient filter. While these users are removed or posting infrequently, by even the most lenient engagement metrics, they contribute with term associations needed to connect dictionary terms in the KG via shortest paths, and are thus kept by our method.

For all social media platforms, very few high engagement users are removed by the backbone approach ($0\%$ for EFA). To understand what type of high-engagement users are removed by the backbone filtering method,
we studied the 1\% of high-engagement Instagram users who were not backbone contributors. Nearly one third had no term co-mentions in any of their posts. 
Of the other two thirds, around half of their co-mentions were rarely used by other users so they did not appear in the KGs  per our inclusion criteria (Section \ref{knowledge network}). 
This shows that even the aggressive engagement filter can select prolific users with no relevance to 
biomedical analysis. In contrast, the backbone method,  keeps only users who co-mention at least one pair of terms that contribute to at least one a shortest path in the KG, regardless of engagement activity.

 
To show that the backbone filtering method helps to identify the most relevant biomedical discourse and users, we used the corpus human-annotated of 1,387 Instagram posts with 2,381 dictionary term occurrences (Section \ref{Sec:InstagramAnnotatedSet}).
Specifically, we evaluated the performance of the filtering methods by checking the proportion of the dictionary terms in a cohort (retained or not retained by filter) that were used with their intended meaning, i.e. with putative biomedical relevance within the context of that specific post.

\begin{figure}
    \centering
    \includegraphics[width=0.75\linewidth]{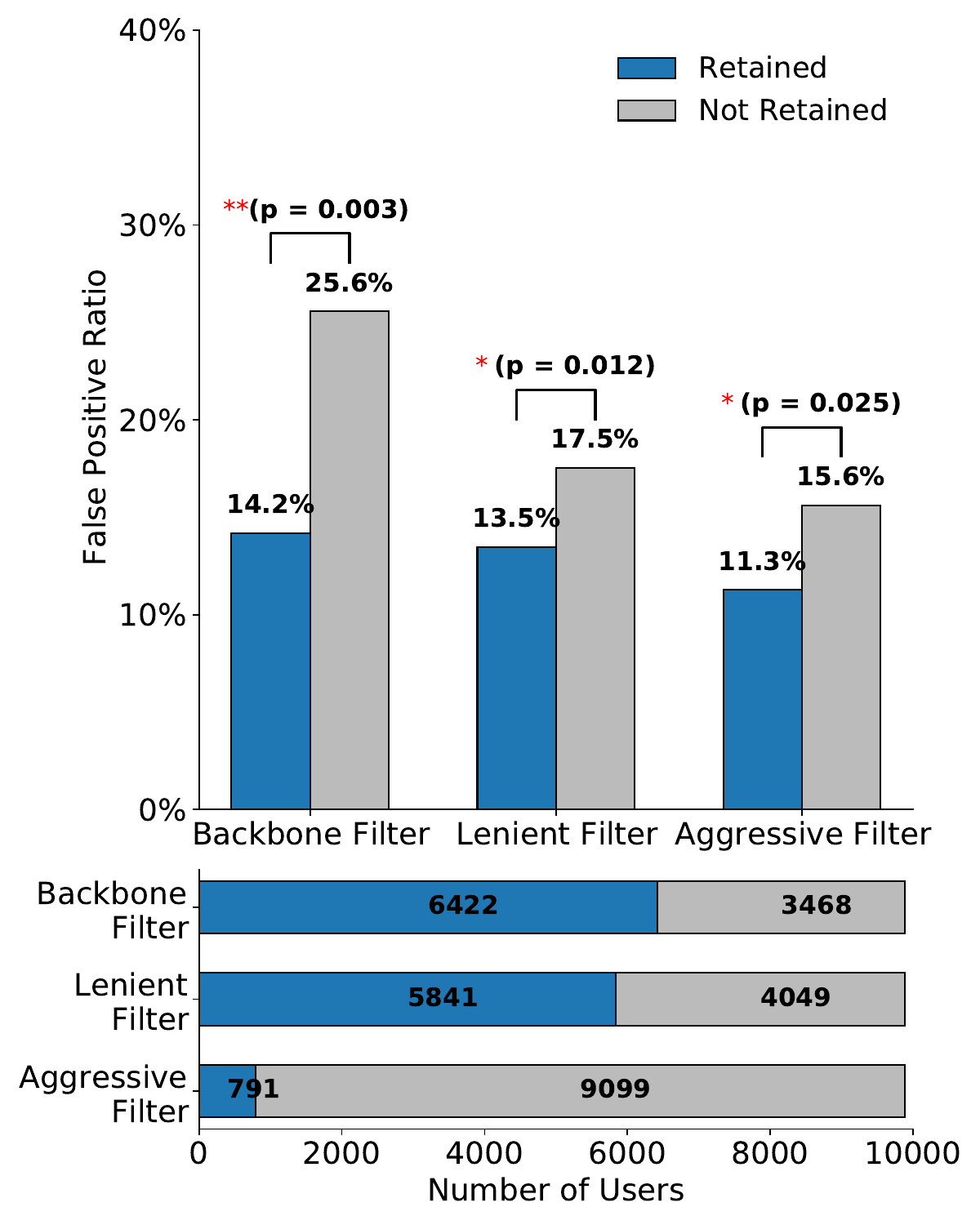}
    \caption{ Study of incorrect term usage per filtering method using a human-annotated corpus of Instagram posts. The \textbf{False Positive Ratio} (FPR) is the proportion of annotated terms from a given cohort (retained or not retained) that are used in contexts unrelated to biomedical inference. The $p\text{-value}$ above each pair of bars indicate the statistical significance of the difference in FPR between retained and not retained users, where \textbf{$**$} indicates $p < 0.01$ and \textbf{$*$} indicates $p < 0.05$. The confidence intervals for the difference in FPR between retained and not retained users are 11.4\% $\pm$ 9.1\%, 4.0\% $\pm$ 3.3\%, 4.3\% $\pm$ 3.3\%, for backbone, lenient, and aggressive filters, respectively, based on a 95\% confidence level. The bottom horizontal bars present the number of retained and not retained users for each filter. 
    }
    \label{fig:ValPlot}
\end{figure}

Figure \ref{fig:ValPlot} depicts the false positive ratio (FPR) for users retained and not retained for the various filtering methods.
Specifically, the FPR is the proportion of false positive terms over all annotated terms in a cohort (see the numbers of annotated term occurrences of each cohort in Table A.10 in Supplementary Material). That is, posts with a dictionary term used not with its intended meaning. For instance ``Valium'' not as a medication, but as a metaphor for something boring \cite{min2024refinement}. 
In the case of our backbone filtering method there is a clear and significant difference between the retained and removed users.
Indeed, the set of not retained cohort has a much higher FPR than the backbone contributors: 25.6\% to 14.2\%, respectively. The 11.4\% 
difference in FPR between the retained and not retained cohort is statistically significant ($p\text{-value} = 0.003$) according to a two proportions $Z$-test, introduced in \ref{Sec:InstagramAnnotatedSet}, which supports the effectiveness of the backbone filtering method in removing irrelevant users from the digital cohort. 
In contrast, traditional engagement filtering methods, including both lenient and aggressive filters, exhibit a considerably lower ability to differentiate between retained and non-retained users with only 4.0\% and 4.3\% absolute difference and $p\text{-value}$ of 0.012 and 0.025 respectively.
More importantly, backbone filtering method allows us to preserve more users, which is approximately 10\% and 700\% more than lenient and aggressive filtering methods, respectively, while removing users with significantly larger FPR. The lenient \textit{use-level} approach (as detailed in Supplementary Material, Section I) has similar result, but the engagement filters do not even show a significant difference in FPR between retained and not retained users.

In short, while the backbone filter removes users who are significantly likely to use biomedical terms accidentally, with incorrect meanings, engagement filters remove users who are much less distinguishable from the retained users in how correctly they use the target biomedical terminology.
These findings support our hypothesis that backbone contributors are more likely to engage in the relevant discourse captured by a target biomedical dictionary.
They also underscore the superiority of our backbone-based approach over conventional engagement-based methods for filtering social media users to obtain focused digital cohorts relevant to biomedical research.

\section{Discussion}
\label{sec:discussion}

%
Discourse on social media contains a wealth of biomedical and health information \cite{correia2020mining}, which is exchanged simultaneously with many other irrelevant topics.
Identifying which users contribute to biomedical discourse that is relevant to a particular, condition, community, or research topic is a difficult challenge. To address this challenge, we devised and demonstrated a new method to extract focused digital cohorts from social media.
The method is based on the algebraically principled, parameter-free metric backbone, used to sparsify KGs obtained from measuring the co-occurrence of (relevant) biomedical terms in posts of social media users---initially identified in a coarser manner (with a wider net, as it were).

The metric backbone preserves important topological characteristics of each KG, such as shortest paths, community structure, and propagation dynamics, even while dramatically reducing their density of edges (section \ref{backbone}). 
Indeed, we showed that the metric backbone of social media KGs is very small (Section \ref{sec:result_backbone}).
Thus, the metric backbone of a KG can be interpreted as containing the strongest chains of association among terms in a given dictionary of terms of biomedical relevance (in our case, a curated medical dictionary used to study epilepsy). Since it contains the association chains that are necessary and sufficient to compute all possible shortest paths, the backbone of a KG can be seen as a minimal, focused kernel of the knowledge that can be inferred (via shortest paths) from a social media user sample and a topical dictionary (of relevant terms).
The selection of focused cohorts, in turn, derives from that sparsified kernel by backtracking the co-occurrence data used to compute the weighted edges of KGs. In other words, by reducing the original (coarser) set of social media users to the set of users who contribute to at least an edge (an association between two terms) in the metric backbone of a KG (Section \ref{methodological result}).

We did not pursue comparisons to other sparsification methods since the backbone sparsification used in our work has already been shown to outperform other state-of-the-art network sparsification approaches—--such as the disparity filter and effective resistance---in prior studies \cite{brattig2023contact, panos2023semi, pereira2023distance}
The backbone sparsification approach can be relaxed to consider a degree of breaking the triangle inequality, resulting in a softer thresholding towards the backbone \cite{panos2023semi}, allowing a less strict filtering of digital cohorts.
This approach can be further expanded by incorporating temporal dynamics, whereby we consider KGs as multilayer networks \cite{Correia_eLife_2024} or by integrating more refined pre-filtering of posts that accounts for semantic context and linguistic nuance using machine learning techniques such as GraphRAG or other large language models.
Additionally, cohort filtering, either via our backbone sparsification approach or machine learning methods, could be enhanced through the inclusion of supplementary metadata, such as demographic attributes, user activity patterns, or self-reported diagnoses, which we aim to explore in future work.
%


KG sparsification yields computational benefits in both analysis and visualization \cite{correia2025myaura}, but the method is bound by the computational complexity of metric backbone computation (Section \ref{backbone}).
However, the KGs derived from social media datasets for biomedical dictionaries contain a manageable number of nodes ($N$, on the order of thousands), which allows us to perform computations efficiently on standard hardware without significant performance issues. Additionally, because our KGs are sparse, we benefit from faster computations than the theoretical worst-case scenario ($\mathcal{O}(N^3)$) by utilizing known optimizations such as efficient priority queue implementations and parallelizing the independent shortest-path computations across multiple processors to improve scalability \cite{kalavri2016shortest}.
Indeed, shortest-path inference on KG is feasible for much larger graphs, such as a 3 million node KG extracted from Wikipedia used for automatic fact-checking \cite{ciampaglia2015computational}. In the biomedical informatics domain, the approach is perfectly feasible even for multilayer, multispecies protein-protein interactions of close to 30,000 genes \cite{Correia_eLife_2024}. Indeed, the bottleneck in the scalability of network analysis methods in the biomedical domain lies in inferring network dynamics, not shortest-path computation \cite{rocha2022feasibility}.

Our biomedical focus in this paper is epilepsy, one of the most common neurological conditions worldwide. So far, social media discourse surrounding epilepsy has been given limited attention by researchers. 
We computed and provide KGs capturing online epilepsy discourse
on four social media platforms: X, Instagram, Reddit (r/Epilepsy), and the EFA forums  (Section \ref{sec:result_backbone}). The KGs are publicly available via our lab's GitHub\footnote{ https://github.com/CASCI-lab/SocialMediaKnowledgeGraphs}.
These networks have already informed the study of health concerns surrounding epilepsy, from pharmacology \cite{correia2016monitoring} and outcomes \cite{wood2022small}, to human-centered design of epilepsy personal libraries\cite{min2023understanding,min2021just}, and will continue to be used by us and others to understand this multifaceted chronic disease \cite{correia2025myaura}.
Importantly, our framework is also general and can easily be extended to study other chronic health conditions. 

To establish the quality of the obtained epilepsy-focused cohorts, we show that the backbone-based approach to social media user filtering is a better alternative than engagement-based approaches. Rather than focusing on how much a user posts (volume), the metric backbone  filtering method
works by removing users that contribute only redundant data for any putative inferences based on shortest-paths on KGs.
This way, it retains low-engagement users if their posts contain necessary term associations (co-mentions), and it can even remove very active users who do not co-mention any term pairs directed linked via the KG metric backbone (see Figure \ref{fig:overlap users} and section \ref{compare filter methods}).
Furthermore, using a corpus of manually annotated Instagram posts, we confirm that users who do not contribute to the backbone have a significantly higher false positive ratio (FPR) for non-biomedical usage of medical terms compared to backbone contributors (25.6\% vs.\ 14.2\%, respectively; $p\text{-value} = 0.003$) (see Figure \ref{fig:ValPlot}). At the same time, the backbone filter allows us to preserve significantly more users compared to engagement filtering methods, while effectively removing users with substantially larger FPR.

Our hypothesis that backbone contributors engage in discourse more closely related to the topic of interest is additionally supported by the fact that focused social media sites have a much higher proportion of backbone contributors than general-purpose social media sites do (Section \ref{result:different sites}). While almost all users (more than $93\%$) who mention epilepsy drugs on epilepsy-specific social media platforms contribute to the backbone, a substantial proportion of users who similarly mention epilepsy drugs on general-purpose social media do not (between $30\%$ and $35\%$). 
It is significant that this is not merely a result of differences in how often a user mentions terms in our medical dictionary. In fact, users of epilepsy-specific social media sites have fewer term matches, on average, than users of general-purpose social media.

Our method is highly generalizable and can be used to identify cohorts of interest from social media across a wide range of medical conditions by utilizing a topic-specific target dictionary of terms.
As a method developed for social media mining, it encounters the limitations and benefits of any approach to analyze such unconventional data in biomedical informatics \cite{correia2020mining}. For instance, there may be insufficient data to obtain statistical validation for conditions that are rare or are associated with stigma on general-purpose platforms \cite{wood2022small,min2023understanding}. 
The data insufficiency was not an issue here for the amount of data reported in Table \ref{tab:data_overview}. Future work will investigate what is the minimal data needed to construct a meaningful knowledge graph and metric backbone.
However, some of these issues can be overcome in more dedicated platforms or those that allow greater anonymity.  
Indeed, studies on opioid discourse on Reddit highlight how this platform enable users to discuss sensitive topics in a less judgmental environment, leading to increased data availability for more an accurate study of such conditions \cite{balsamo2021patterns}.

Our own work here, focusing on epilepsy which is known to carry much stigma \cite{min2023understanding,correia2025myaura}, suggests that the discourse on the Epilepsy-specific social media platforms Reddit(r/Epilepsy) and EFA forums is more relevant to study the condition than the (pre-filtered) discourse on X and Instagram. Indeed, we show that a much higher proportion of users of the epilepsy-specific sites are relevant for inference than the users of the general-purpose sites, even when pre-selected by mentioning drugs used to treat epilepsy (Section \ref{result:different sites}). Thus, a focus on dedicated and anonymous platforms can lead to sufficient data collection even for conditions that are rare or carry stigma. Notably, more conventional data sources (e.g. electronic health records) are also problematic in studying those conditions, as patients may not be open about their issues even with medical professionals \cite{min2023understanding, correia2025myaura}. 
For instance, patients with epilepsy sometimes do not disclose seizures to their doctors to avoid losing their driver licenses, but will discuss their seizures anonymously on platforms like Reddit and EFA \cite{correia2020mining, mahmud2017understanding, dahiya2024digital}.
Thus, social media mining may still be one of the best alternatives to collect data to study them.

Our approach is ultimately designed to aid biomedical researchers and practitioners who seek to utilize any type of social media data for a specific health-related study---including those with stigma such as epilepsy.
By filtering out irrelevant users, our method provides a more pertinent cohort for analysis.
Indeed, the metric backbone filter allowed us to obtain from X and Instagram, focused digital cohorts that are much more similar to the cohorts extracted from EFA forums and the r/Epilepsy Reddit community---in terms of contributing to the KG backbone (section \ref{result:different sites}) and intended use of relevant terminology (section \ref{compare filter methods}). In situations where such topic-specific social media sites are not available, data from general-purpose social media (like X), which have very large user populations, can thus be filtered to obtain more focused digital cohorts.


The backbone filtering method depends on a specific biomedical dictionary to construct the KGs (Section \ref{knowledge network}). Our dictionaries are constructed from biomedical ontologies such as MedDRA (Section \ref{dictionary}). But a pertinent question is whether using our KG backbone sparsification offers a benefit over using an existing ontology directly to select users for digital cohorts? Our work responds affirmatively to this question in at least two ways. 
First, by building platform-specific KGs, we weight the importance of relationships among terms from a universal ontology, to community- and condition-specific discourse. Indeed, the weighted KGs derived for each social media platform are different, as seen in Figure \ref{fig:NetworkOverview}. We can regard these network representations as a projection of a universal ontology to the particular discourse of specific community---in our case, epilepsy-related discourse on distinct social media platforms.
Second, the derived KGs are amenable to sparsification via the metric backbone that is only applicable to weighted graphs. This allows us to remove redundant associations and users, which here we have shown to lead to more focused, higher-quality digital cohorts.
Indeed, we can compare our results to the alternative approach that would use the dictionary (built from universal ontologies) directly---which would lead to the raw digital cohorts of Table \ref{tab:data_overview}. Thus, our results demonstrate that building and sparsifying the platform-specific KGs allows us to remove many redundant users, deriving digital cohorts much more focused on a condition-specific community of users, than what is obtained from using an ontology directly.

In short, our approach leverages social media data to weight universal (ontology) terminology and their associations to extract digital cohorts focused on a condition of interest.
In future work, one can certainly consider other biomedical ontologies beyond the ones we used in our dictionary, if more appropriate to study specific problems. We hope that our paper motivates other groups to explore different ontologies and dictionaries, to study other conditions on social media or other data sources associated with populations of interest.

\section{Acknowledgement}
Research reported in this paper was supported by the National Library of Medicine of the National Institutes of Health under award \# R01LM012832. The content is solely the responsibility of the authors and does not necessarily represent the official views of the funding agency.
We are thankful to Deborah Rocha for very thorough line editing.

\bibliographystyle{elsarticle-num} 
\bibliography{refs}


\clearpage
\appendix

\end{document}